\documentclass[a4paper,11pt]{article}
\pdfoutput=1 

\usepackage{jinstpub} 

\title{\boldmath Monte Carlo simulations of the S-shaped neutron guide}

\author[a,1]{L.P. de Oliveira,\note{Corresponding author.}}
\author[a]{A.P.S. Souza,}
\author[b]{F. Yokaichiya,}
\author[a]{F.A. Genezini,}
\author[a]{M.K.K.D. Franco}

\affiliation[a]{Projeto do Reator Multiprop\'{o}sito Brasileiro,\\Instituto de Pesquisas Energ\'{e}ticas e Nucleares (IPEN), S\~{a}o Paulo - Brazil}
\affiliation[b]{Department Quantum Phenomena in Novel Materials,\\Helmholtz-Zentrum Berlin f\"{u}r Materialen und Energie (HZB), Berlin - Germany}

\emailAdd{luiz.oliveira@ipen.br}

\abstract{Neutron transport along guides is governed by the Liouville theorem and the technology involved has advanced in recent decades. Computer simulations have proven to be useful tools in the design and conception of neutron guide systems in facilities. In this study, we use a Monte Carlo method to perform simulations for an S-shaped neutron guide with different dimensions for a Small-Angle Neutron Scattering (SANS) instrument, through the MCSTAS software. A wavelength cutoff is observed and shown to be dependent on the geometrical parameters of the guide. Results for the neutron flux at sample position are presented and a greater sensitivity of cutoffs concerning the curvatures of the guides than to their lengths is noticed. Our results are in agreement with those obtained from the Acceptance Diagram method and we analyze the beam divergence behavior along the S-shaped guide.}

\keywords{Instrumentation for neutron sources, Neutron sources, Instrument optimization, Simulation methods and programs.}

\begin{document}
\maketitle
\flushbottom

\section{Introduction}
\label{sec:intro}

The project of neutron facilities and instruments are usually complex and expensive. The development of detailed study of the neutron instrumentation necessarily requires a previous description of the neutron beam properties. Monte Carlo (MC) simulation is an essential tool for defining instrument components and their operational limitations (e.g., the wavelength operational range) as much as guide geometrical definition, which implies in wavelength profile at instrument entrance. MCSTAS is a software that uses the MC method to simulate neutron scattering for various configurations, including different sources, samples and components \cite{lefmann1996}.

The neutron transportation is ruled by the Liouville theorem \cite{liouville1838}, which states that the neutron phase-space volume of moments and positions is conserved. Therefore, the optimization of facility neutrons transport systems and instruments, frequently achieved with MC simulations, is necessary to obtain a fine resolution and high intensity, which provide high performance to neutron instruments. In the last two decades, neutron engineering has been improved in the optimization of large nuclear installations \cite{romain2016}. In this scenario, there are many studies in literature that analyse the relation between the neutron guides, flux intensity, divergence and wavelength resolution at facilities instruments \cite{artus2000,wang2002}.


In the optimization of neutrons guides, there are some basic distinctions between straight and curved guides that are related to neutron transmission based on their wavelength. According to literature is observed an asymmetry of neutron distribution along with the curved guide, in opposition to straight guide. The geometry of a curved guide is crucial for determining the way neutrons transported along guide by multiple reflections in its inner walls, which are divided into two possible regimes, namely Garland and Zigzag. Here, the Garland regime consists of neutron hitting just outer guide mirror reflections and the Zigzag regime, in opposition, described neutron transportation considering reflections in both walls. In these terms, both regimes of reflection are responsible for bringing asymmetries in neutron flux distributions as much as the change of initial wavelength profile distribution (that enter in the guide system) \cite{maierleibnitz1963,mildner1990}.

A well-established way to theoretically study the neutron flux behavior conducted by a curved guide is through the Acceptance Diagrams (AD) \cite{carpenter1982,mildner1990}. In this study is proposed to investigate the relation of curved guides dimensions and the intensity, spatial and angular distribution of transmitted neutrons by employing AD. After this analysis, distribution results are related to the number of reflections according to each regime (Garland or Zigzag) that the transported neutron flux belongs.

From curvature ($R$) and width ($w$) of the guide, we can predict that neutrons are transmitted more efficiently and also study specific neutron trajectory inside the guide. Curved guide systems are generally designed to avoid the direct length of sight, which mandatorily excludes unwanted neutrons and gamma to fly directly to the instrument entrance position.  For this purpose, curved guides possess a minimum length given by $L_{ds}=\sqrt{8Rw}$ \cite{mildner1990}. Nevertheless, there are also studies for AD of the called ``short'' guides, where their length is less than $L_{ds}$, where other guide segments connected after system allow avoiding the direct sight \cite{copley1995}.

According literature \cite{romain2016,maierleibnitz1963,mildner1990,copley1995}, we observe that the definition of a guide system is crucial for fulfilling the demand for a specific flux profile at each neutron instrument. In this context, the reallocation of instruments could be a puzzle situation from a engineering point of view. This occurred at the Forschungs-Neutronenquelle Heinz Maier-Leibnitz (FRM-II, Germany) that received eight instruments of FRJ-2 reactor from the Jülich (Germany), which was decommissioned by the time. Besides the limited area for new instrument installation, the FRM-II team also dealt with a $1.2$ $m$ displacement between instruments and beam hole planes. To solve these problems, the scientific facility team started studying the application of the named ``S-Shaped Guide'', which consists of a curved guide followed by another similar curved guide in the opposite sense (forming a guide shaped like a letter ``S'') \cite{radulescu2008}. 

Additional studies that exploit properties of these S-shaped guides have also been done. By using MC simulations, Gilles and collaborators were able to determine a nominal wavelength cutoff for neutrons transmitted through guides \cite{gilles2006}. This property has shown itself a useful way to reduce instrument background in addition to displace guide entrance and exit planes (vertically or horizontally). There is still inside AD theory a study of neutron behaviour passing through S-shaped guides \cite{mildner1990}, where the authors found a wavelength cutoff utilizing its geometry and correspondent diagram. The concept of a S-shaped guide was also applied in the SAFARI reactor in South Africa in a way to circumvent the natural asymmetry that a curved guide imposes to the neutron flux \cite{hofmeyr1974,hofmeyr1979}.

We believe that a better understanding of S-shaped guide could open news possibilities of instrument installations in operating facilities and the redistribution of some instruments to other ones. Recently, the Helmholtz Zentrum Berlin (HZB) Supervisory Board decided for shutting down and dismantling the German reactor BER-II until the end of 2019.  From this decommissioning came the agreement between ANSTO and HZB for the installation of the reflectometer SPATZ at Australian reactor OPAL \cite{lebrun2016}, which is an example of instrumentation adaptation where beforehand planning and studies are necessary.  In this sense, we also observe a problem similar to FRM-II in a previous examination of SANS installation at the Brazilian reactor IEA-R1, where there is a space limitation for allocating this instrument close to beam holes. Thus, an alternative to alleviate this problem could be to displace SANS to the upper floor with an S-shaped guide.

Considering the properties of wavelength cutoff and the geometrical versatility in adapting instruments in real facilities, as describe above, the motivation of this study is to investigate neutron transportation through different S-shaped guide configurations by using MCSTAS simulations. In this paper we study the dependence of wavelength cutoff and S-shaped guide curvature and length, once this cutoff is independent on the initial flux specter (facility source) \cite{luiz2019}. Moreover, we study the influence of collimator length on the final detected flux divergence as much as the correlation between Mildner cutoff \cite{mildner1990} and simulated wavelength cutoff. This paper is organized as follows: Section~\ref{SSGMS} contains a basic configuration of an S-shaped guide and the MCSTAS simulation description; Section~\ref{RAD} describes simulation results and respective discussions; Section~\ref{C} shows our final conclusions and some future perspectives of application and studies.

\section{S-shaped guide and MCSTAS simulations}
\label{SSGMS}
 
The S-shaped guide provides a vertical displacement $z$ of position of the sample in relation to the beam hole, ideal for different sample environments. The dependence of vertical displacement ($z$) as a function of the radius ($R_{i}$) and lengths ($L_{i}$) of S-shaped guide parts are presented below

\begin{equation}
z=\sum_{i=1}^{2}R_{i}\left [ 1 - \cos\left ( \frac{L_{i}}{R_{i}} \right )  \right ].   
\end{equation}

We can observe that in the limit $L_{i}/R_{i}$ $\ll$ 1, the vertical displacement takes the form $z~=~{L_{1}^2}/{2R_{1}}~+~{L_{2}^2}/{2R_{2}}$, which provides a zero displacement for a straight guide ($R$~$\longrightarrow$~$\infty$). 

\begin{figure}[b!]
    \centering
    \includegraphics[scale = 0.18]{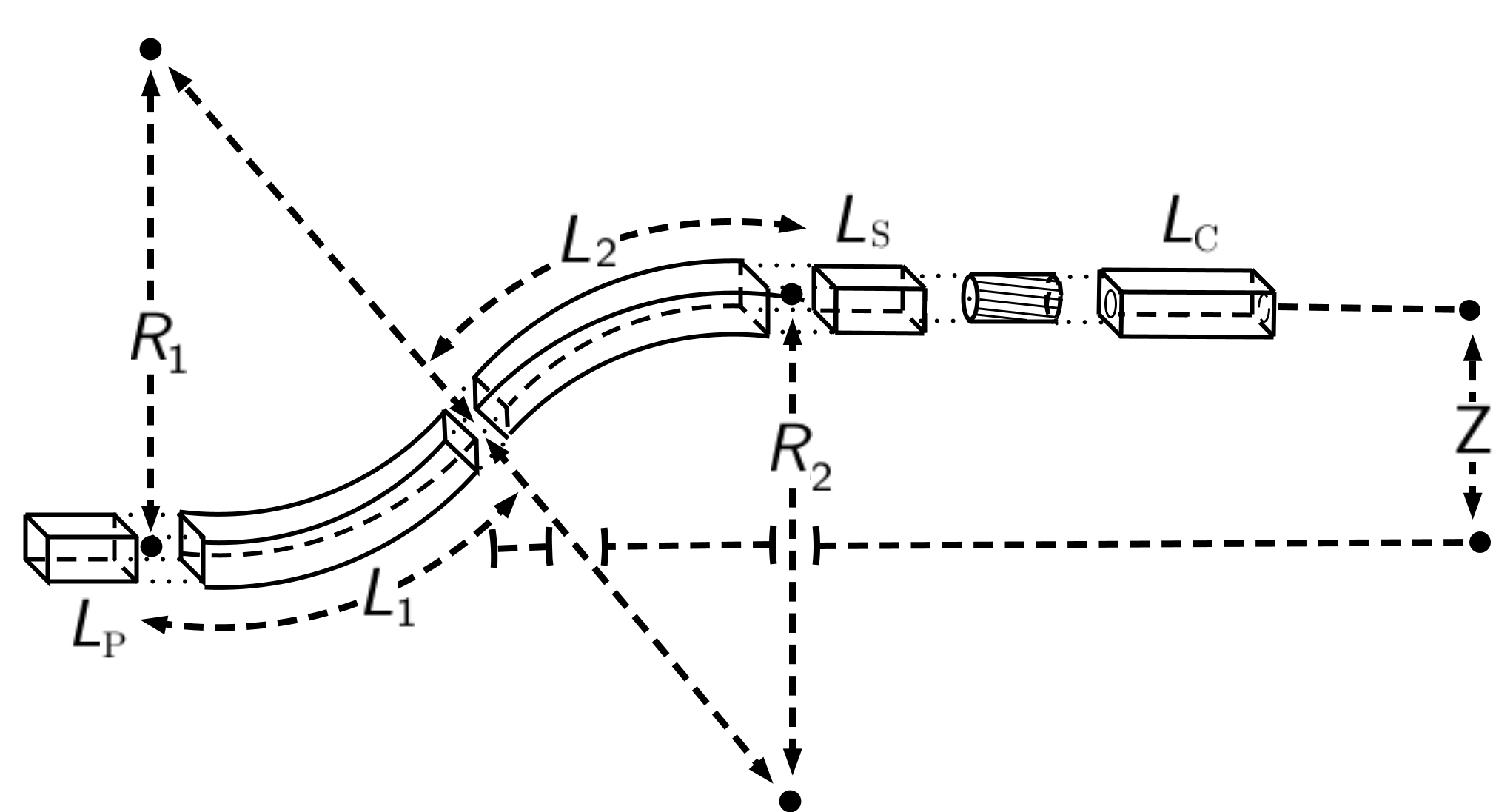}
    \caption{Side-view sketch of S-shaped guide. The system is sequentially composed of a primary guide with length $L_P$, two curved guides with length and curvature of $L_1$, $L_2$ and $R_1$,$R_2$, respectively, a secondary guide with length $L_S$, a velocity selector and a collimator with length $L_C$. The S-shaped guide system provides a vertical displacement of $z$.}
    \label{s_shaped_guide_sketch}
\end{figure}

\begin{table}[]
\centering
\caption{\label{tab:i} Parameters of MCSTAS simulations. All values of $L_S$, $L_1$, $L_2$, $R_1$, $R_2$, and $L_C$ used in simulations are shown.}
\smallskip
\begin{tabular}{|l|c|}
\hline
\textbf{Simulation Parameters}           & \textbf{Values}                               \\ \hline
Source Area (Radius)                     & $15.5$ $cm$                                     \\
Source-Guide Distance                    & $7.45$ $m$                                      \\
Guide Areas (Width $\times$ Height)      & $5 \times 5$ $cm^2$                             \\
Primary Guide Length ($L_P$)                     & $1$ $m$                                       \\
Secondary Guide Length ($L_S$)                  & $1$, $2$ and $20$ $m$                           \\
S-shaped guides Length ($L_1$ and $L_2$)         & $8$, $12$, $16$, $20$, $24$ and $32$  $m$        \\
S-shaped guides Curvature ($R_1$ and $R_2$)     & $75$, $125$, $250$, $500$, $750$ and $1000$ $m$ \\
Velocity Selector Length                 & $25$ $cm$                                     \\
Velocity Selector Rotation Frequency     & $141.5$ $Hz$                                  \\
Velocity Selector Twist Angle            & $48.3^\circ$                                        \\
Velocity Selector Chose Wavelength                      & $ 15$ \AA                                  \\
Collimator Length ($L_C$)                       & $1$, $2$ and $20$ $m$                               \\
Collimator Slits size (Radius)           & $1$ $cm$                                        \\
Supermirror ($m$)                        & $2$                                           \\
Number of shot neutrons (Ray Count)        & $10^7$ neutrons                                       \\ \hline
\end{tabular}
\end{table}

All carried out MCSTAS simulations are based on Figure~\ref{s_shaped_guide_sketch} basic sketch. Here, the studied system is composed of a straight primary guide with a fixed length ($L_P$). This guide is followed by a vertical curved guide connected to another curved guide in the opposite sense, forming themselves an S-shaped guide. Their lengths and curvatures are $L_1$ and $L_2$ and $R_1$ and $R_2$ as also shown in Figure~\ref{s_shaped_guide_sketch}. This S-shaped guide is plugged into a straight secondary guide with variable length ($L_S$), which is in sequence followed by a collimation system composed of a velocity selector and a collimator.  The S-shaped guide can be symmetric ($R_{1}~=~R_{2}$ and $L_{1}~=~L_{2}$) or asymmetric ($R_{1}~\neq~R_{2}$ and $L_{1}~\neq~L_{2}$; $R_{1}~=~R_{2}$ and $L_{1}~\neq~L_{2}$; $R_{1}~\neq~R_{2}$ and $L_{1}~=~L_{2}$). Additionally, gravity is not considered in our simulations and it is physically expected that the results are invariant by the exchange $R_{1}$ and $L_{1}$ by $R_{2}$ and $L_{2}$, respectively, since they provide in the same value of $z$.

This work presents five different types of simulation (1-5), each one of them with distinct cases. To study the geometric dependence of the system, we have chose different cases of simulation with $R_1~=~R_2=~250$~$m$ and $L_1~=~L_2=16$~$m$, independently. For the first case (Simulation 1), we have carried out simulations with six values of guide curvatures, namely  $R_1~=~R_2~=~75$, $125$, $250$, $500$, $750$ and $1000$~$m$ with fixed lengths of $16$~$m$. For the Simulation 2, we have that simulations with a fixed $R_1~=~R_2~=~250$~$m$ have been shot with six values of guide length, namely $L_1~=~L_2~=~8$, $12$, $16$, $20$, $24$ and $32$~$m$. To verify the intensity and distribution of neutron flux at the sample place of a virtual Small-Angle Neutron Scattering instrument (SANS) (Simulation 3), we have simulated a fixed scenario of an S-shaped guide with $R_1~=~R_2~=~250$~$m$ and $L_1~=~L_2~=~16$~$m$ with different values of collimator length, i.e., $L_C~=~1$, $2$ and $20$~$m$ (with fixed $L_S~=~1$~$m$). The role of the velocity selector in this simulation is also taken into account in Simulation 3.
To check the distribution behavior of neutron flux inside the S-shaped guide, we perform Simulation 4, which consists on repeating Simulation 3 and obtaining position distribution results of neutron flux at the entrance (before primary guide), in the middle (between curved guides with $R_1~=~R_2~=~250$~$m$,  $L_1~=~L_2~=~16$~$m$) and ate the end of the system (after secondary guide with $L_S=1$~$m$).

The last set of simulations (Simulation 5) consisted of investigating how a straight guide after the S-shaped guide contributes to a homogeneous vertical distribution intensity at instrument entrance. This case have been carried out by fixing the instrument configuration of $R_1~=~R_2~=~250$~$m$,  $L_1~=~L_2~=~16$~$m$ and $L_C~=~1$~$m$ for a variable secondary guide length with three different values, namely $L_S~=~1$, $2$ and $20$~$m$ . 

All performed simulations (1--5) have been carried out with supermirror guides with $m~=~2$. Besides, simulations (3--5) are modeled with basic dimensions $R_1~=~R_2~=~250$~$m$ and $L_1~=~L_2~=~16$~$m$ that provide an operational value of $z~=~1$~$m$, which correspond to the same order of displacement applied to FRM-II instrument adaptation \cite{radulescu2008}. 

The MCSTAS components \verb|Source_gen()|, \verb|Guide()|, \verb|Guide_curved()|, \verb|V_selector()| and \verb|slit()| form, in sequence, the main structure of simulations. Reference systems (\verb|Arm()|) and monitors (\verb|Monitor_nD()|) are allocated all over the code to respectively relate components and verify flux behavior along with guides. The first component (\verb|Source_gen()|) has been modeled to emulate the FRM-II cold source utilizing Maxwellian parameters intensity ($I_1$, $I_2$ and $I_3$) and temperature ($T_1$, $T_2$ and $T_3$) \cite{mcstas}. Straight guides are described with \verb|Guide()| and the S-shaped guide is constructed with two \verb|Guide_curved()| components. The most important flux measurements are performed with a detector before the first guide (S-shaped guide entrance), in the connection of both curved guides (in the middle of S-shaped guide), between the last guide and velocity selector (S-shaped guide exit) and after the collimator, which is constructed with two slits and one guide in simulations. Parameters and specifications of MCSTAS components are described in Table~\ref{tab:i}.

\section{Results and discussions}
\label{RAD}

\begin{figure}[h!]
    \centering
    \includegraphics[scale = 0.55]{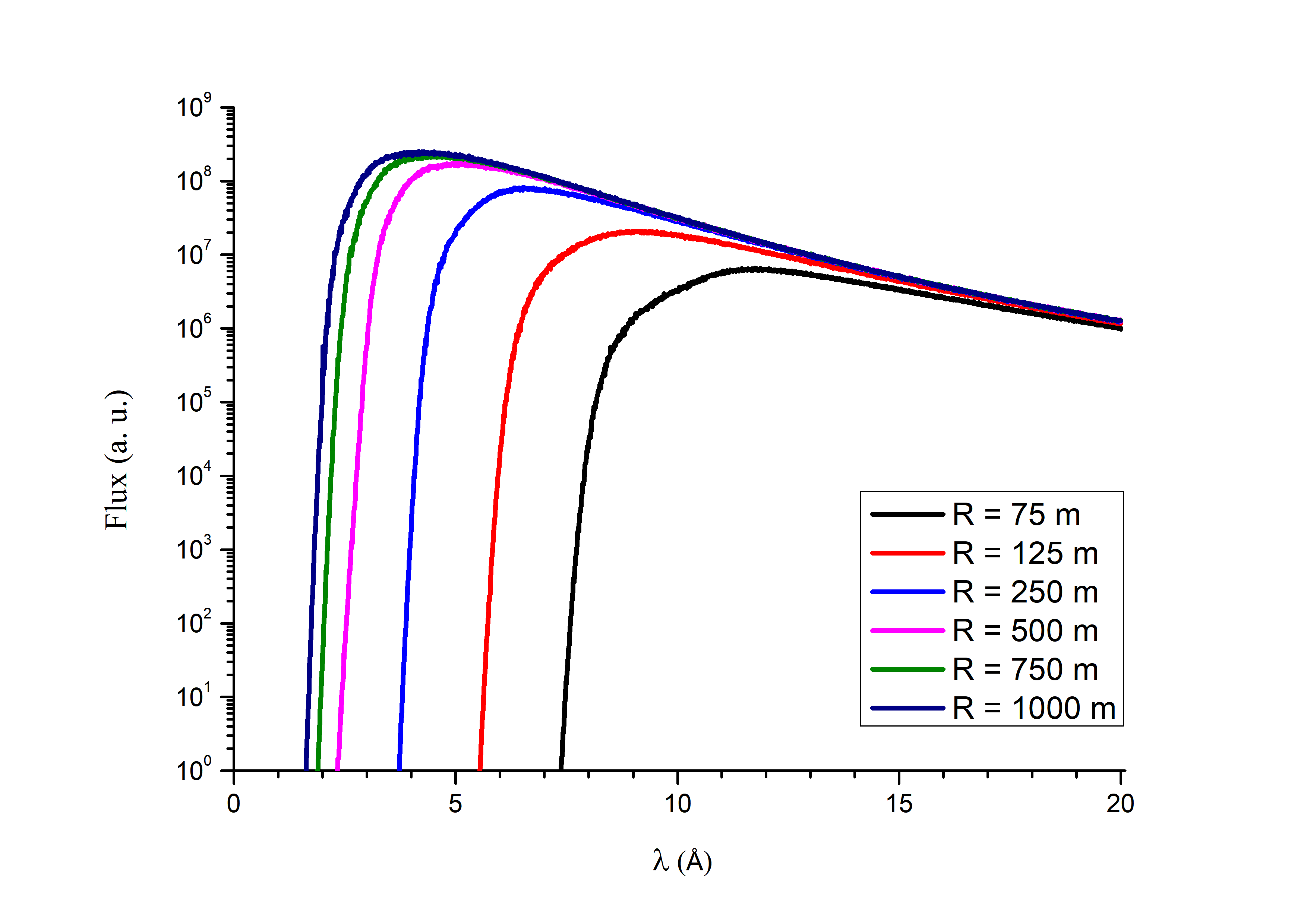}
    \caption{Simulations for a S-shaped guide with $L_{1}~=~L_{2}~=~16$~$m$. Neutron profile (intensity versus wavelength) relations are obtained for different values of curvature in symmetric disposal ($R_1~=~R_2~=~R$). The black curve stands for $R~=~75$~$m$, the red one for $R~=~125$~$m$, the  blue for $R~=~250$~$m$, the purple for $R~=~500$~$m$, the green for $R~=~750$~$m$ and the dark blue for $R~=~1000$~$m$.}
    \label{L=16m_profile}
\end{figure}

\begin{figure}
    \centering
    \includegraphics[scale = 0.55]{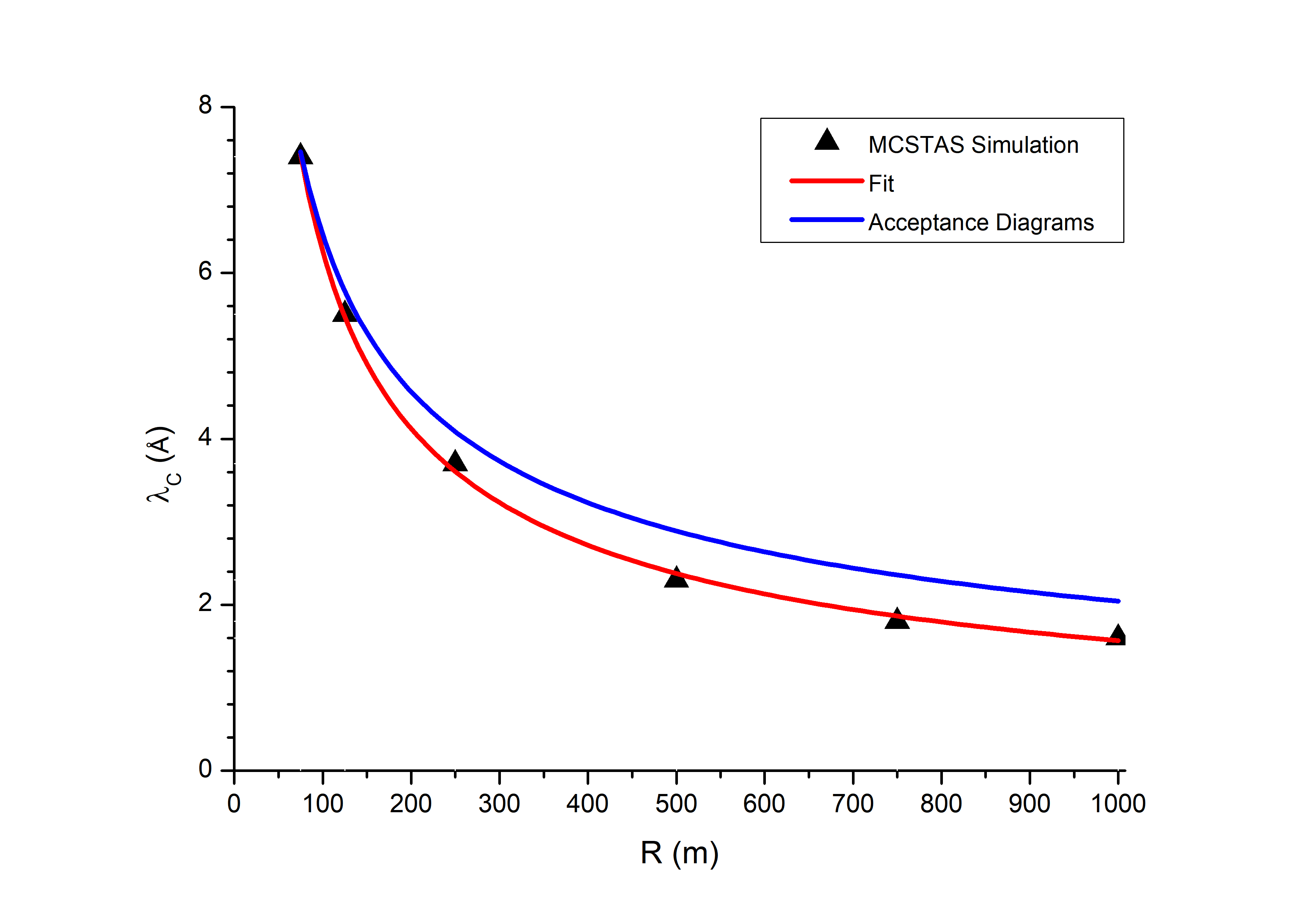}
    \caption{Wavelength cutoff in function of R with $L_{1}~=~L_{2}~=~16$~$m$ for two distinct sources. The black triangles represent MCSTAS simulation data with a red line curve fitting them. The blue curve stands for an analytical result that comes from the Acceptance Diagram formalism.}   \label{theo_vs_sim}
\end{figure}

The parameters used in our simulations (see Table~\ref{tab:i}) provide values of $z$ found between $0.25$ $m$ and $39$~$m$. 
Figure~\ref{L=16m_profile}, which corresponds to Simulation 1 results, shows a decrease of $\lambda_{c}$, according to the increase of curvature of the guides, as well as the increase of the peak of the flux, i.e., more neutrons are absorbed (transmitted) as $R$ decreases (increase). The dependence of wavelength cutoff is shown in Figure~\ref{theo_vs_sim}. From data, we have been able to fit a curve with square correlation factor of $0.9991$.  In addition, this curve is given by  $\lambda_{c} (R)~=~102\times~R^{-0.6060}$, where $\lambda_{c}~\longrightarrow~0$ to $R~ \longrightarrow~\infty$, i.e., straight guides allow the passage of neutrons from all accessible energies.

\begin{figure}
    \centering
    \includegraphics[scale = 0.55]{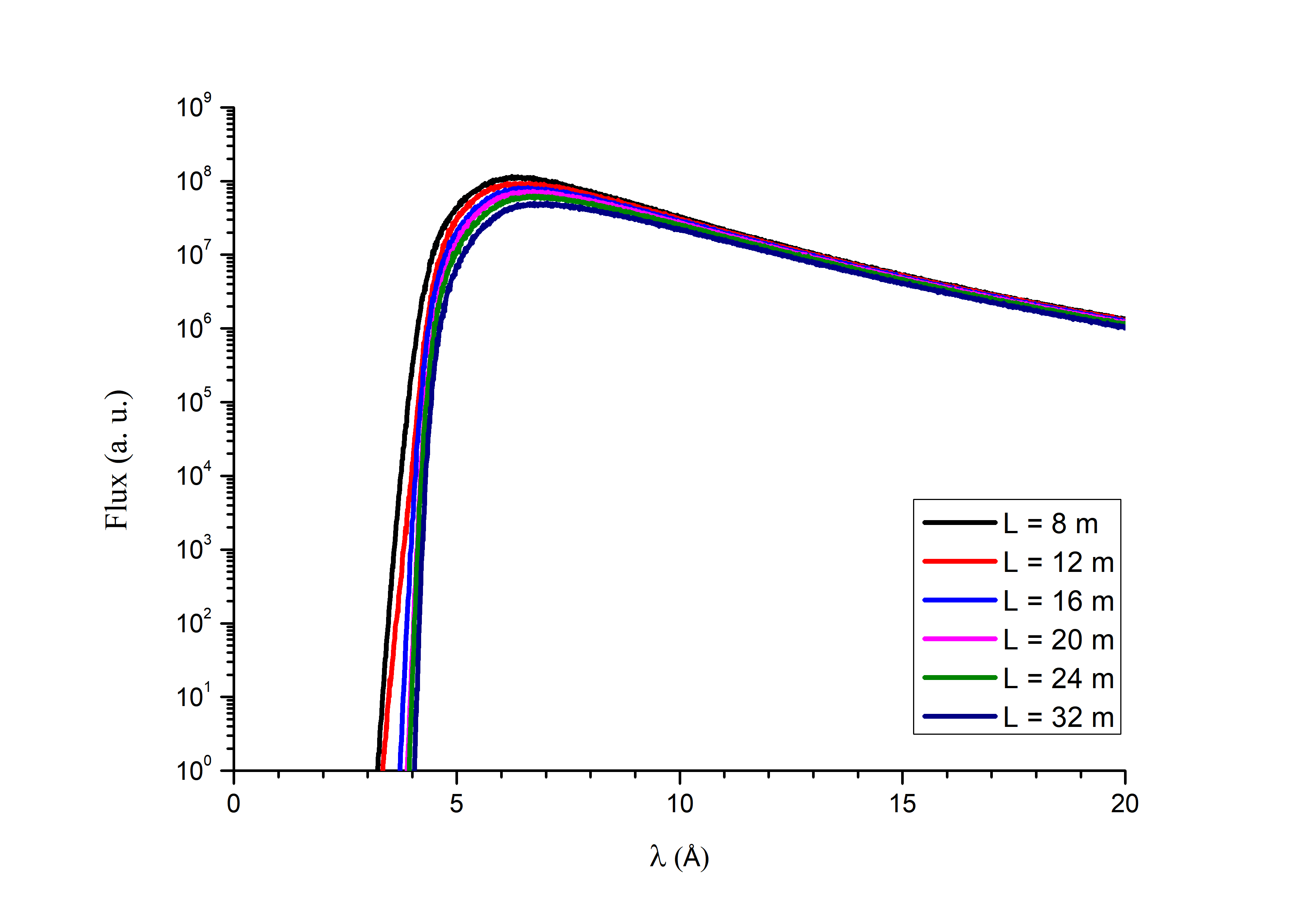}
    \caption{Simulations for a S-shaped guide with $R_{1}~=~R_{2}~=~250$~$m$. Neutron profile (intensity versus wavelength) relations are obtained for different values of guide length in symmetric disposal ($L_1~=~L_2~=~L$). The black curve stands for $L~=~8$~$m$, the red one for $L~=~12$~$m$, the  blue for $L~=~16$~$m$, the purple for $L~=~20$~$m$, the green for $L~=~24$~$m$ and the dark blue for $L~=~32$~$m$.}
   \label{R=250m_profile}
\end{figure}

\begin{figure}
    \centering
    \includegraphics[scale = 0.51]{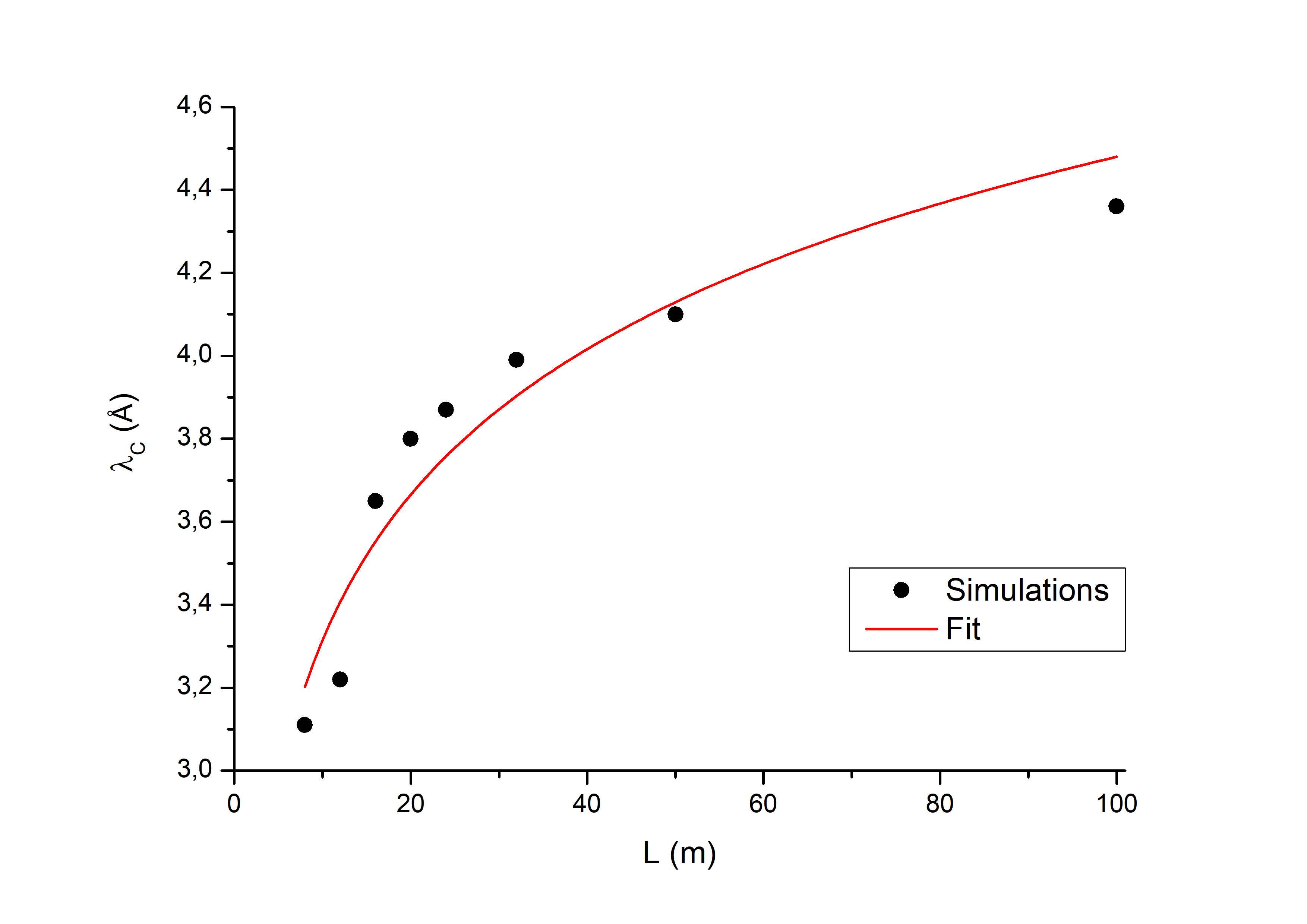}
    \caption{Wavelength cutoff in function of L with $R_{1}~=~R_{2}~=~250$~$m$. The black points represent wavelength cutoff data from simulations presented in Figure~\ref{R=250m_profile} with two extra points obtained from additional simulations for $L~=~50$ and $100$~$m$. The red line curve represents the fit of these data.}
    \label{peaks_profile}
\end{figure}

%

The convergence of all curves in Figure~\ref{L=16m_profile} in $\lambda~=~14$~\r{A} shows an expected physical result since the slower neutron conduction has a very efficient transport, not being directly affected by the S-shaped guide. Figure~\ref{theo_vs_sim} shows the behavior of cutoff wavelength (triangles) as a function of $R$. The red curve is a fit of the simulation results, while the blue curve represents the results extracted from analytical study of Mildner \cite{mildner1990} through AD. We can observe a good agreement between the two studies for smaller $R$ values, while both curves diverge as much as $R$ increases. The percentage deviation between the two approaches can be found in Table~\ref{tab:ii}, which shows a maximum divergence of $24.11\%$. The criteria to define the cutoff wavelength is based on picking a $\lambda_c$ value with correspondent neutron flux less than $0.1$~$a.u.$. We verified that our results could show better convergence to the results of the analytical study if taken in neutron fluxes of the order of $10^3$~$a.u.$, i.e., much larger than our criteria. 

\begin{table}[]
\centering
\caption{\label{tab:ii} Theoretical and simulated values of wavelength cutoff for cases with $R_1~=~R_2~=~R$ and S-shaped guide length of $16$~$m$ (i.e., $L_1~=~L_2~=~16$~$m$). $L^{*}$ stands for the characteristic guide length, which avoids the direct sight of a curved guide. The last column presents the difference between wavelength values. }
\smallskip
\begin{tabular}{|c|c|c|c|c|}
\hline
\text{$R$ $(m)$} & \text{$L^{*}$ $(m)$} & \text{\begin{tabular}[c]{@{}c@{}}$\lambda_c$ (\AA)\cite{mildner1990}\\ Theoretical\end{tabular}} & \text{\begin{tabular}[c]{@{}c@{}}$\lambda_c$ (\AA)\\ Simulations\end{tabular}} & \text{$\Delta\lambda$ $(\%)$} \\ \hline
75                 & 5.48               & 7.46                                                                                            & 7.45                                                                                            & 0.13                            \\
125                & 7.07               & 5.78                                                                                            & 5.47                                                                                            & 5.39                            \\
250                & 10.00              & 4.09                                                                                            & 3.59                                                                                            & 12.10                           \\
500                & 14.14              & 2.89                                                                                            & 2.36                                                                                            & 18.32                           \\
750                & 17.32              & 2.36                                                                                            & 1.85                                                                                            & 21.76                           \\
1000               & 20.00              & 2.04                                                                                            & 1.56                                                                                            & 24.11                           \\ \hline
\end{tabular}
\end{table}

By considering the criteria to state a wavelength cutoff, we have verified the difference between intensities at theoretical cutoff values and the peak intensity of distribution in simulation results. In other words, we have checked how large is the intensity of theoretical wavelength cutoff next to the highest intensity value in profile (peak). Results show that a difference between $0.1\%$ for $R~=~1000$~$m$ and $0.002\%$ for $R~=~75$~$m$. In these terms, we could consider a good agreement between simulation and the Mildner results. This kind of result is already expected since associated errors in MCSTAS simulations with ray count $10^7$ could roughly vary between $0.1\%$ and $10\%$ depending on the wavelength detector bin. 

\begin{figure}[h]
    \centering
    \includegraphics[scale = 0.4]{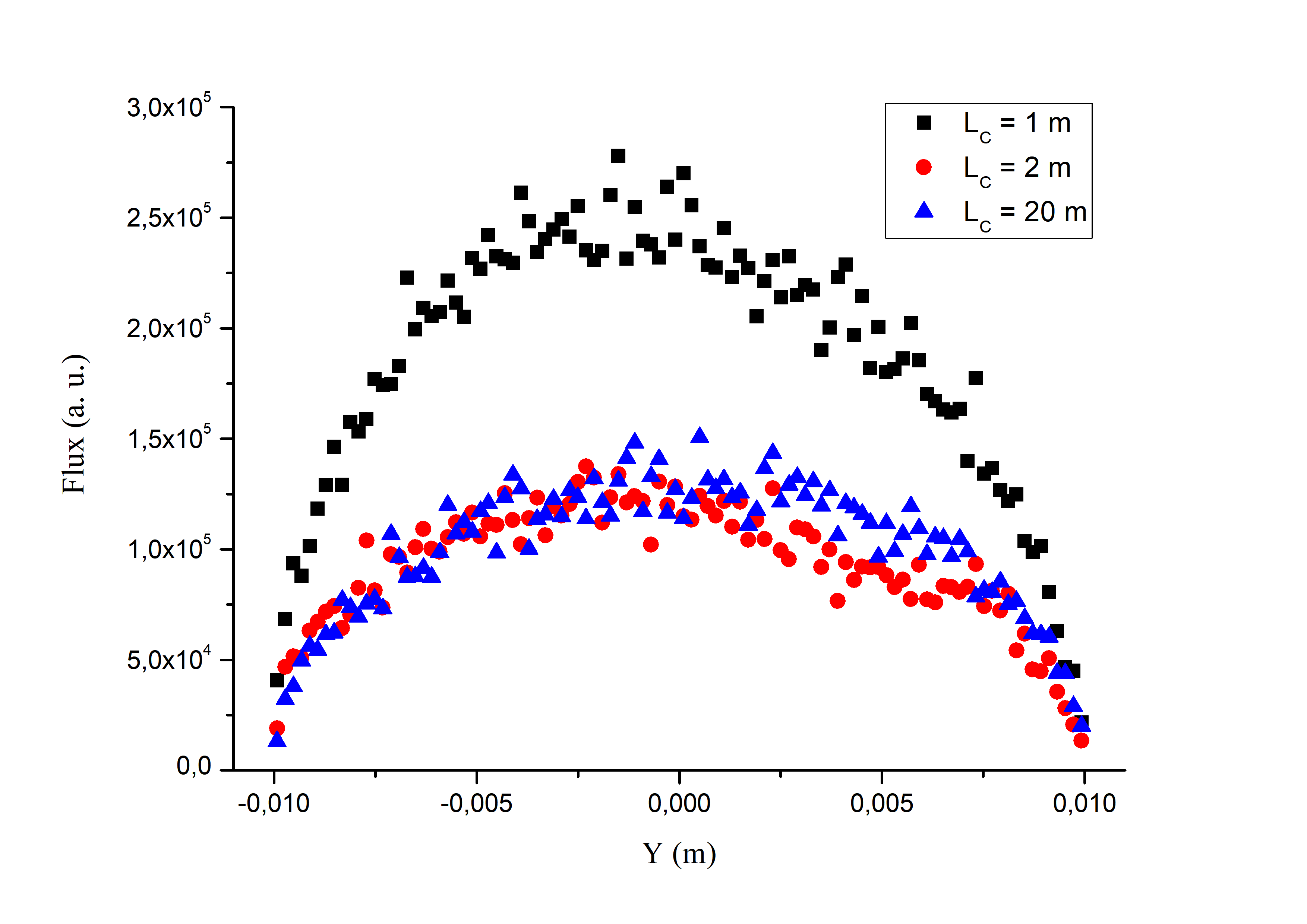}
    \caption{Vertical distribution of neutron intensity at the sample place (after collimator) for $L_{c}~=~1$, $2$ and $20$~$m$, $L_{1}~=~L_{2}~=~16$ $m$ and $R_{1}~=~R_{2}~=~250$~$m$. The black squares, red circles and blue triangles stand for MCSTAS simulation data results for collimation length ($L_C$) of $1$, $2$ and $20$~$m$, respectively.}
   \label{flux_sample}
\end{figure}

The dependence of the neutron flux with the length of the guides is found in Figure~\ref{R=250m_profile} (Simulation 2). A weak dependence of the wavelength cutoff with the length of the guides ($\lambda_{c}$ = 3--4 \r{A}) is observed for $R_{1}~=~R_{2}~=~250$~$m$ (Figure~\ref{peaks_profile}). We have also been able, with additional simulations with guides length longer than $32$~$m$, to fit the data by an adjusted equation $\lambda_{c}$~(L)~$=$~$ln{(L^{0.51})}~+~2.15$, where we find a small cutoff wavelength variation between $3.1$~\r{A} and $4.4$~\r{A} (square correlation factor of $0.9162$).

\begin{figure}[!htb]
    \centering
    \includegraphics[scale = 0.13]{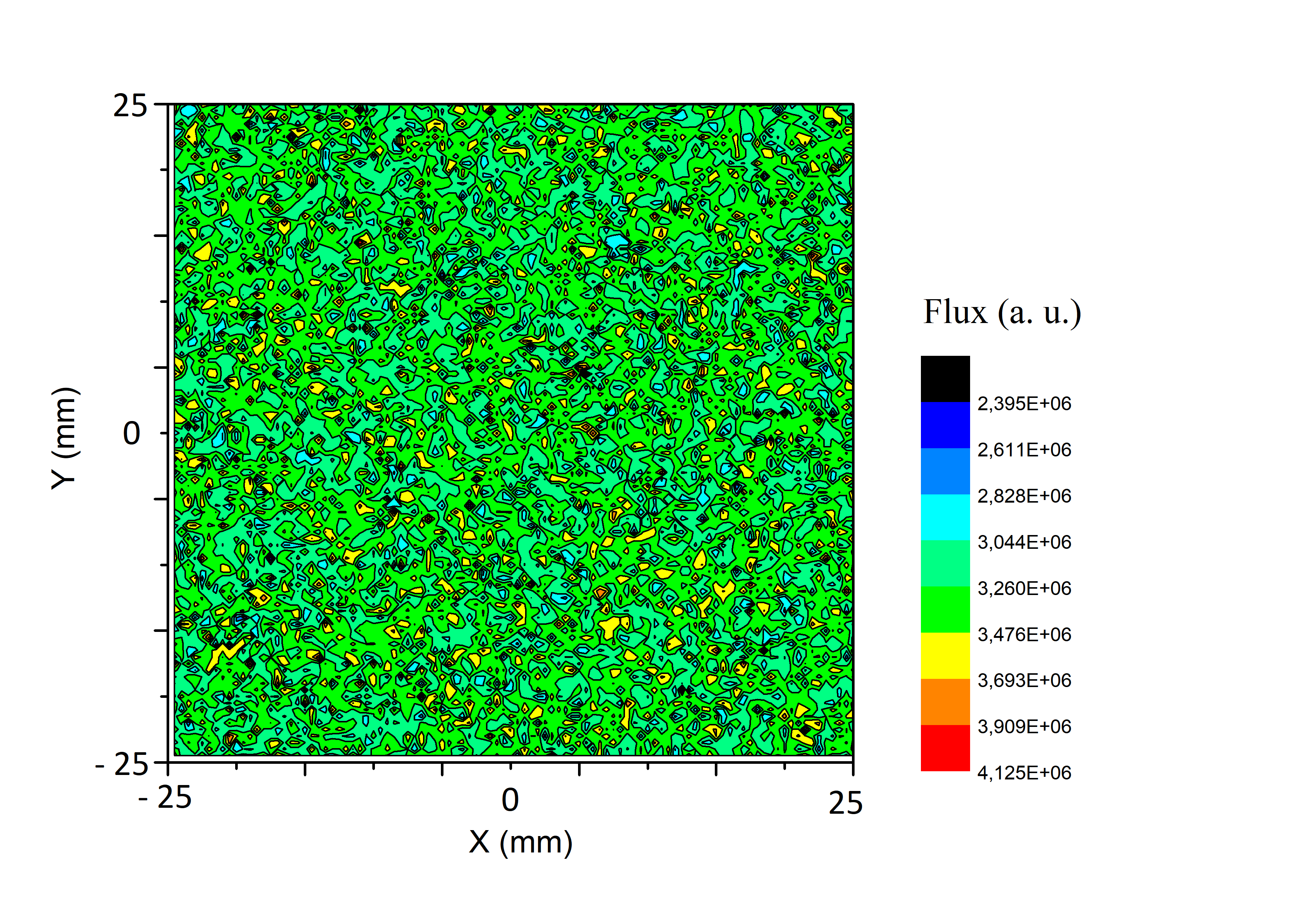}
    \caption{Neutron position results (X versus Y) of a MCSTAS virtual detector allocated at the entrance of an S-shaped guide (before primary guide) with $R_1~=~R_2~=~250$~$m$ and $L_1~=~L_2~=~16$~$m$ (standard).``Hotter'' colors, tending to red, represent greater densities of neutron detection into the figure color scale.}
    \label{PDS_ENTRANCE}
\end{figure}

\begin{figure}[!htb]
    \centering
    \includegraphics[scale = 0.13]{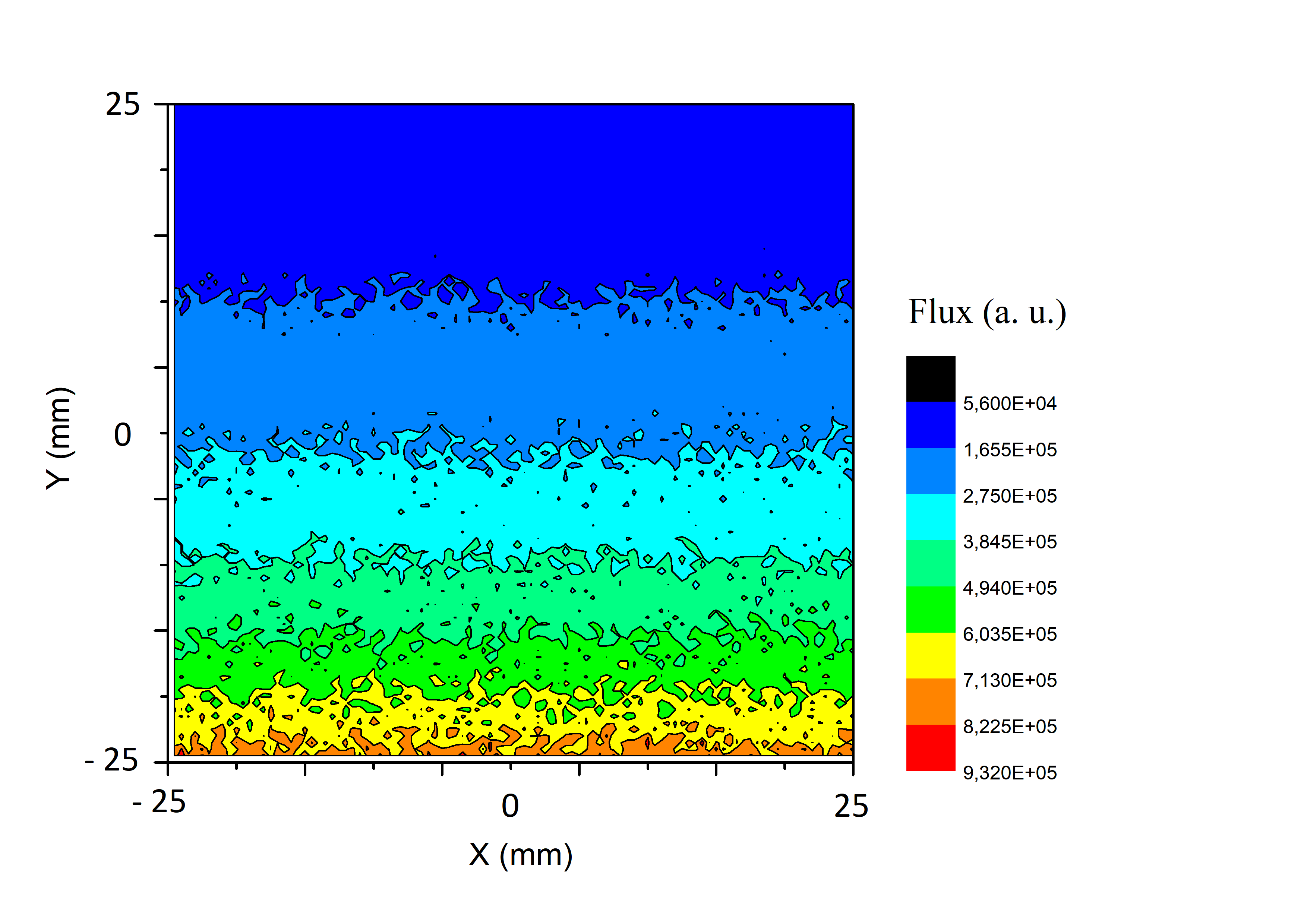}
    \caption{Neutron position results (X versus Y) of a MCSTAS virtual detector allocated in the middle of an S-shaped guide (between two curved guides) with $R_1~=~R_2~=~250$~$m$ and $L_1~=~L_2~=~16$~$m$ (standard).``Hotter'' colors, tending to red, represent greater densities of neutron detection into the figure color scale.}
    \label{PDS_MIDDLE}
\end{figure}

\begin{figure}[!htb]
    \centering
    \includegraphics[scale = 0.13]{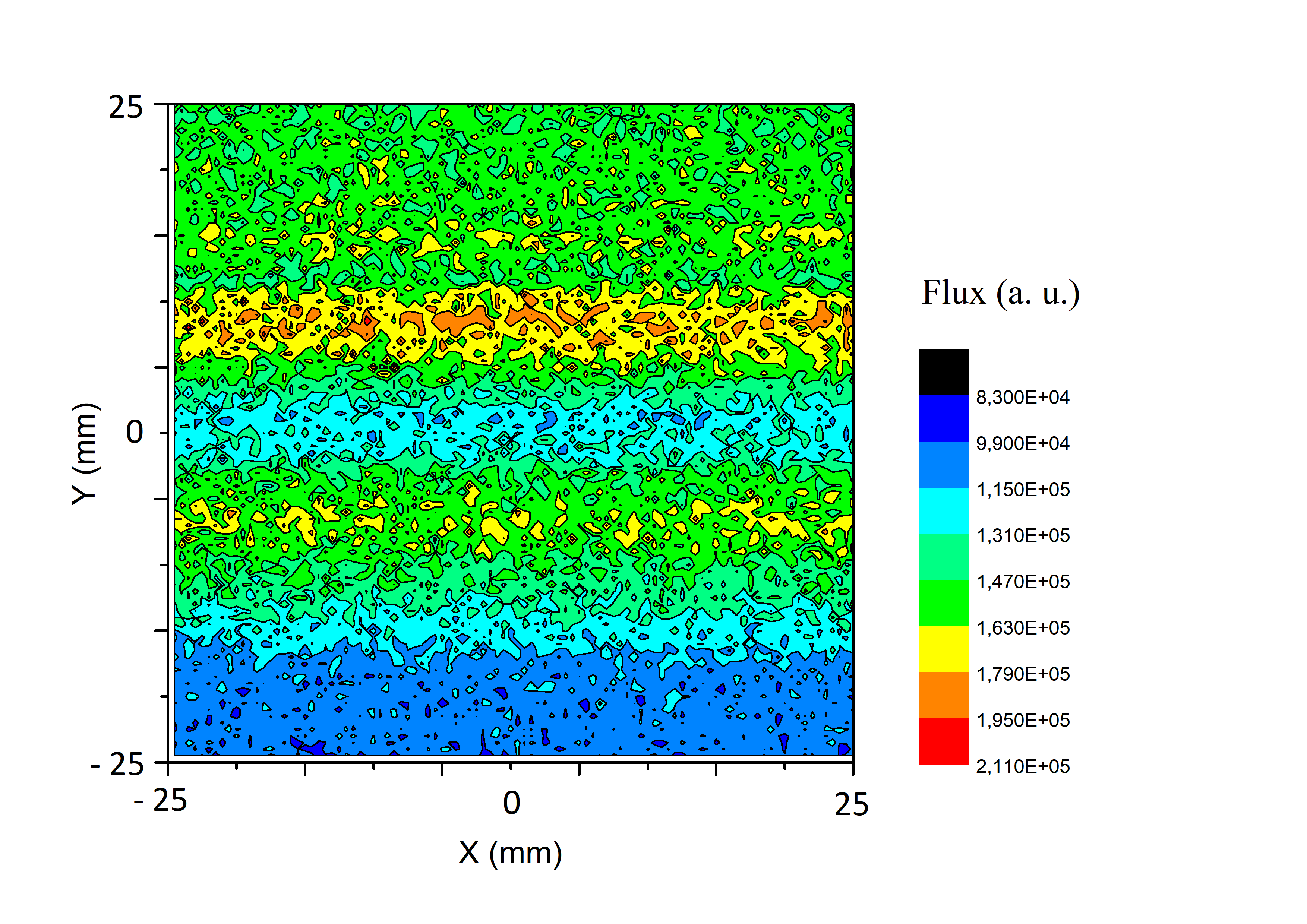}
    \caption{Neutron position results (X versus Y) of a MCSTAS virtual detector allocated at the end of an S-shaped guide (after secondary guide) with $R_1~=~R_2~=~250$~$m$ and $L_1~=~L_2~=~16$~$m$ (standard).``Hotter'' colors, tending to red, represent greater densities of neutron detection into the figure color scale.}
    \label{PDS_END}
\end{figure}

AD method reveals a dependence between the cutoff wavelength and the radius of curvature, but shows independence of the guide length. Therefore, there is no comparative analysis between our results and the paper of Mildner\cite{mildner1990}. According to AD theory, it is given that the only relation between wavelength cutoff and S-shaped guide length is the necessity to avoid direct sight from the neutron source and instrument entrance once this is the main purpose of any curved guide. Taking this into account, we can state that values of characteristic guide length ($L^{*}$) are minimal limits to guarantee wavelength cutoff. In other words, simulated cases of Table~\ref{tab:ii} could have $L_1~=~L_2~=~L^{*}$ instead of $L_1~=~L_2~=~16$~$m$, which in some cases could shorten guides and increase flux at their exits\footnote{Values of $L^{*}$ are avoiding the direct sight for each curved guide of an S-shaped guide individually (i.e., $L_1~>~L^{*}$ and $L_2~>~L^{*}$). In this sense, we probably verify the cutoff process for $L^{*}~>~16$~$m$ because the direct sight of the whole S-shaped guide is guaranteed.  However, additional simulations have to be performed to better understand the relation between the cutoff and guide length.}. The results for asymmetric S-shaped guides have curves with same behavior shown in Figures~\ref{L=16m_profile} and \ref{R=250m_profile}, locating in the adopted intervals $75~m<R<1000~m$ and $8~m<L<32~m$.

Figure~\ref{flux_sample}, correspondent to Simulation 3 results, shows the flux along with the vertical position at the sample place for different values of collimator length ($L_C$), namely $1$, $2$ and $20$~$m$ for a guide of $R_1~=~R_2~=~250$~$m$ and $L_1~=~L_2~=~16$~$m$, which is the standard case. From these results, we observe that there is no significant variation in detected flux for longer values of collimation length than $2$~$m$. Such results are possibly explained by the fact that the velocity selector has already filtered well-collimated neutrons, for cases where $L_C~>~2$~$m$, that pass through both collimator slits without reflecting in its internal coat.   

According to literature and the study of AD, there is a natural inhomogeneous distribution of neutrons in the guide section depending on its curvature.  This process, which depends on neutron wavelength, also generates in inhomogeneous patterns of neutron distribution in S-shaped guides since it is built by two curved guides. By checking Figures~\ref{PDS_ENTRANCE}~-~\ref{PDS_END}, we observe the neutron distribution behavior in three different points of a standard S-shaped guide case utilizing MCSTAS position detectors (Simulation 4). These three detectors have been allocated at the entrance  (Figure~\ref{PDS_ENTRANCE}), in the middle (Figure~\ref{PDS_MIDDLE}) and at the end of the S-shaped guide (Figure~\ref{PDS_END}).

\begin{figure}[h!]
    \centering
    \includegraphics[scale = 0.4]{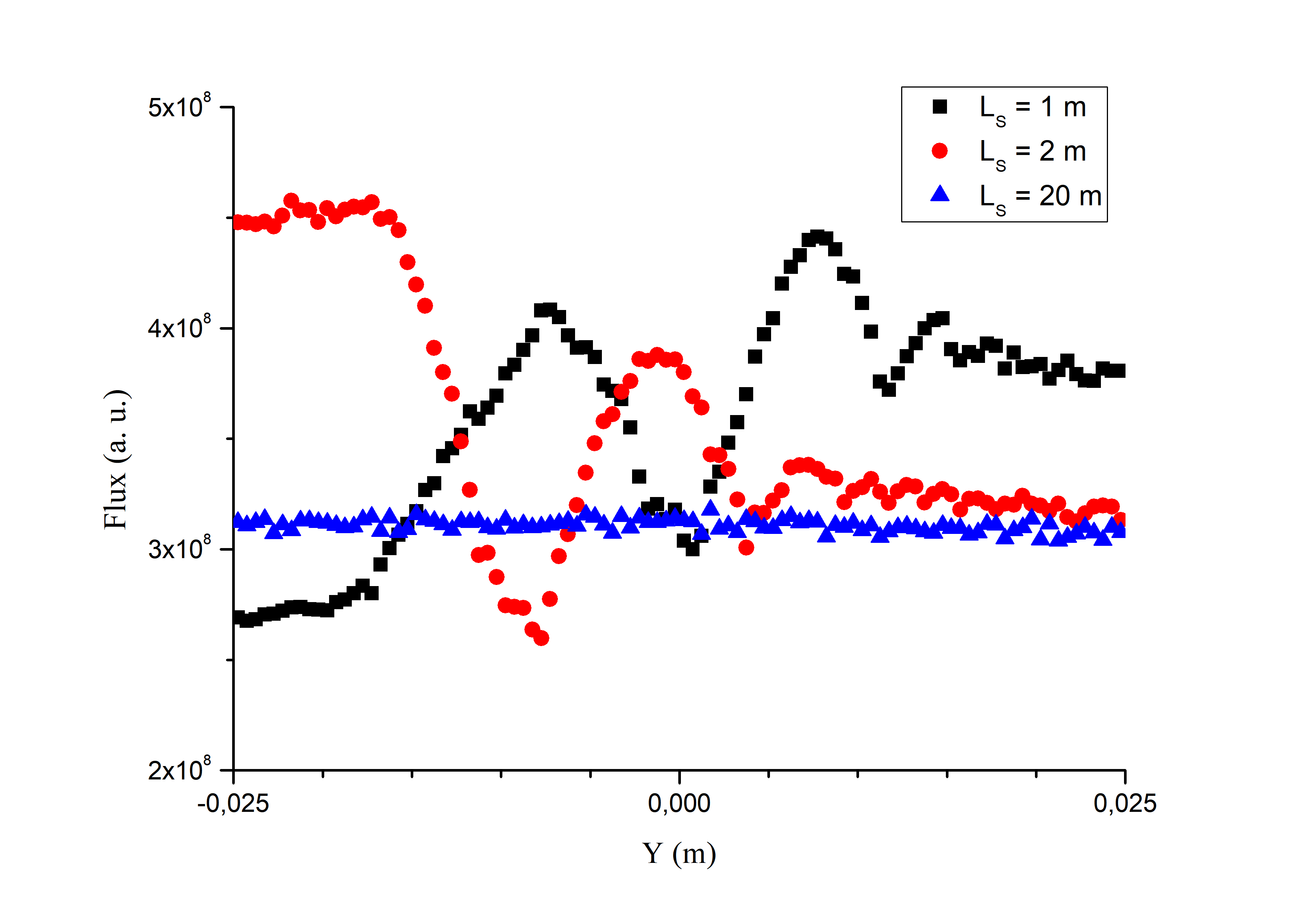}
    \caption{Vertical  distribution of neutron intensity at the end of S-shaped guide (after secondary guide) for $L_{S}~=~1$, $2$ and $20$~$m$, $L_{1}~=~L_{2}~=~16$~$m$ and $R_{1}~=~R_{2}~=~250$~$m$.  The black squares, red circles and blue triangles stand for MCSTAS simulation data results for secondary guide length ($L_S$) of $1$, $2$ and $20$~$m$, respectively.}
    \label{flux_s_shaped_guide}
\end{figure}

At the entrance of the S-shaped guide, we observe a homogeneous distribution of neutron all over the detector (Figure~\ref{PDS_ENTRANCE}). After being transmitted along with the first curved guide, neutrons distribution displace a larger quantity of neutron to the outer side of the guide (which is at the bottom of the Figure \ref{PDS_MIDDLE}) according to the Garland reflection regime. On the other hand, the last detector, which is at the end of the S-shaped guide, shows an inhomogeneous distribution, but symmetrically better-distributed concerning the detector center. Besides, we observe that angular divergence naturally follows the same inhomogeneous behavior, which implies that a homogeneous neutron distribution is desired at the instrument entrance. Simulations have shown us that most of the transmitted neutrons are confined into a divergence ``area'' of $\Delta\theta_h~=~2^\circ$ per $\Delta\theta_v~=~2^\circ$, where subscribed letters ``h'' and ``v'' stand for horizontal and vertical angular divergence, respectively. However, the curvature of the S-shaped guide imposes asymmetries in distribution inside this ``area'',  which has to be turned to homogeneous patterns before entering into the instrument.

To circumvent both divergence and neutron distribution problems, some authors analyzed different methods to smooth neutron distributions after a simple curved guide for instance. In publications \cite{hofmeyr1974,hofmeyr1979}, the authors used another curved guide and obtain, consequently an S-shaped guide. However, it is common to proceed by utilizing a straight guide segment to turn neutron distribution homogeneous even for curved or S-shaped guides. In this sense, by considering that neutron flux of the S-shaped guide configuration does not result in a homogeneous vertical distribution of neutrons intensity, we have tested three different secondary guides length to investigate distribution behavior. Our simulations on this matter consist of verifying the vertical distribution intensity relation at the end of the S-shaped guide for three different length values of the secondary guide (Simulation 5). Simulations have been carried out with $L_S$ of $1$, $2$ and $20$~$m$  and also for the S-shaped guide standard case (i.e., $R_1~=~R_2~=~250$~$m$ and $L_1~=~L_2~=~16$~$m$).

Simulation results are presented in Figure~\ref{flux_s_shaped_guide}, where black squares, red circles and blue triangles stand for flux values for $L_S~=~1$, $2$ and $20$~$m$ , respectively.  From such results, we observe that it is necessary larger values than $L_S=2$~$m$ for imposing a homogeneous distribution after an S-shaped guide system. We also observe in Figure~\ref{flux_s_shaped_guide} that there is no significant loss of flux for distributions with $L_S=1$ and $2$~$m$, and for $L_S=20$~$m$ (i.e., with same order of magnitude). In this sense,  a long collimation system (if there is available space at the facility) could be applied without a high cost of intensity. 

It is worth to note that each configuration of the S-shaped guide posses its own values of collimation to guarantee a low divergence and homogeneous flux distribution. In this scenario, individual simulations are then necessary to find the proper secondary guide length. By comparing Figures~\ref{flux_sample} and \ref{flux_s_shaped_guide}, it is possible to check that a velocity selector plays an important role in divergence and distribution homogeneity since all distribution in Figure~\ref{flux_sample} have the same ``regular'' shape, differently than Figure~\ref{flux_s_shaped_guide}.

\section{Conclusions}
\label{C}
The use of an S-shaped guide is an available possibility in instrument adaptation and installation at reactor facilities. In this study, we explore other aspects of S-shaped guides related to their geometry and the transmitted neutron flux. Even under rigid criteria of wavelength cutoff determination in simulation, we observe that there is a good agreement between simulation results and analytical values obtained via AD. We are also able to verify a weak relation between S-shaped guides length and wavelength cutoff, differently from its dependence on guides curvature, which is proportional to $R^{-0.5}$ and $R^{-0.6060}$ for the analytical and simulation approach, respectively.

Results for different collimation lengths (i.e., for different $L_C$ and $L_S$) are also considered. Simulations indicate a small dependence between collimator length and intensity (and angular divergence) for collimators longer than $2$~$m$. By comparing these results with simulations with different secondary guide lengths, we check that the velocity selector plays an important influence in divergence and neutron position distribution results. This happens since only the case of $L_S~=~20$~$m$ demonstrates a homogeneous angular divergence and position distribution in opposition to collimator simulations, where all three results attend these properties. These results accordance with literature and reassure the need of utilizing straight guide after curved ones to improve system outcoming flux. 

Our results show that the geometrical aspect of the S-shaped guide is a more important achievement than the wavelength cutoff process itself. For modeling the instrument incoming neutron flux profile, other optical components, like filters and benders, would be desirable. In this scenario, there is current and initial speculation of the installation of a SANS instrument at Brazilian reactor IEA-R1.  Nevertheless, a lack of space at the reactor neutron hall opened a possibility of displacing instrument installation from ground stage through an S-shaped guide. The vertical displacement ($z$) between $3$~$m$ and $4$~$m$ could provide a neutron flux at the upper floor of the neutron instrument hall, however a detailed study of flux wavelength cutoff and profile next to S-shaped guide geometry is planned for future works. 

\acknowledgments

APSS and LPO would like to thank CNPq for financial support under grant numbers 381565/2018-1 and 380183/2019-6, respectively.
\bibliographystyle{JHEP2}
\bibliography{bibliography}










\end{document}